\RequirePackage{silence}
\documentclass[]{aa}

\usepackage[varg]{txfonts}

\usepackage{graphicx}

\NewCommandCopy{\AALogoOld}{\AALogo}
\renewcommand*{\AALogo}{\href{https://www.aanda.org}{\AALogoOld}}

\DeclareRobustCommand{\ion}[2]{%
  \textup{#1\,\raisebox{0pt}[1ex][0pt]{\scalebox{0.85}{\uppercase{#2}}}}%
}

\makeatletter
\idline{\ifnum\value{page}=1{\myFullArticleNumber\hspace{3.5cm}\textcolor{red}{arXiv version, typsetting different from A\&A}}\else{%
\aa@authorrunning: \myFullArticleNumber}\fi}
\renewcommand*\aa@copyrightname{\copyright~The Authors}
\renewcommand\aa@manuscriptname{%
  manuscript no. \aa@numarticle
  \hspace{\stretch{1}}%
  \aa@copyrightname \the\year
}
\renewcommand\aa@textidlineempty{{\slshape A\&A proofs:}\ manuscript no.~\aa@numarticle}
\renewcommand*\doi[1]{%
  \renewcommand*\aa@doi{\href{https://doi.org/#1}{https://doi.org/#1}}%
  \renewcommand*\aa@doifig{#1}%
}
\IfFileExists{latexml.sty}{
  \usepackage{latexml} 
}{\newif\iflatexml\latexmlfalse}
\iflatexml\else\fancypagestyle{firstpage}{%
    \fancyhf{}%
    \renewcommand*{\headrulewidth}{\z@}%
    \renewcommand*{\footrulewidth}{\z@}%
    \fancyfoot[RO]{\aa@footfont \aa@numarticle\aa@pageof}%
    \fancyfoot[LE]{\aa@footfont \aa@numarticle\aa@pageof}%
    \fancyfoot[LO]{\\[0.5\baselineskip]\includegraphics[height=16pt]{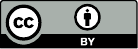}}
    \fancyfoot[C]{\\[0.5\baselineskip]\scalebox{0.78}{%
            \begin{tabular}{@{}c@{}}%
                \tiny Open Access article, published on \href{https://www.arxiv.org}{arXiv}, under the terms of the Creative Commons Attribution License (\href{https://creativecommons.org/licenses/by/4.0}{https://creativecommons.org/licenses/by/4.0}),\\
                \tiny which permits unrestricted use, distribution, and reproduction in any medium, provided the original work is properly cited.
            \end{tabular}%
        }%
    }%
}\fi
\newcommand{\latexMLheader}{\myFullArticleNumber\hfill\href{https://www.aanda.org}{Astronomy \& Astrophysics}\\\aa@doi\\\copyright~The Authors \the\year}
\makeatother
\newcommand{\hereOrThere}[2]{\iflatexml\providecommand{#1}{#2}
\else\providecommand*{#1}{}#2\fi}
\iflatexml\renewcommand{\tablefoot}[1]{#1}\else\fi
\AddToHook{shipout/lastpage}{\label{LastPage}}

\usepackage{hyperref}
\hypersetup{
    colorlinks=true,
    allcolors=blue
    }

\newcommand*{\myFullArticleNumber}{A\&A, 706, A218 (2026)}
\AANum{A218}
\doi{10.1051/0004-6361/202558127}

\begin{document} 
\makeatletter
\renewcommand*\aa@pageof{, page \thepage{} of \pageref{LastPage}} 
\makeatother

\iflatexml\latexMLheader\else\fi

   \title{The magnetic origin of the outer boundaries of sunspots}


   \author{ Markus Schmassmann\inst{1}$^,$\thanks{Corresponding author: \href{mailto:schmassmann@leibniz-kis.de}{\texttt{schmassmann@leibniz-kis.de}}}\href{https://orcid.org/0000-0002-7303-1006}{\includegraphics[height=8pt]{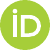}}, 
              Nazaret Bello González\inst{1}\href{https://orcid.org/0000-0002-0479-9134}{\includegraphics[height=8pt]{ORCID-iD_icon_vector.pdf}}, 
            Jan Jurčák\inst{2}\href{https://orcid.org/0000-0002-9220-4515}{\includegraphics[height=8pt]{ORCID-iD_icon_vector.pdf}}, and
            Rolf Schlichenmaier\inst{1}\href{https://orcid.org/0000-0002-7386-8578}{\includegraphics[height=8pt]{ORCID-iD_icon_vector.pdf}}}

   \institute{  
             Institut für Sonnenphysik (KIS), Georges-Köhler-Allee 401a, 79110 Freiburg, Germany
             \and
         Astronomical Institute of the Czech Academy of Sciences, Fričova 298, 25165 Ondřejov, Czech Republic}

   \date{Received 14 November 2025 / Accepted 15 December 2025}

 
  \abstract
   {Sunspot boundaries are commonly outlined by contours of the continuum intensity. However, their magnetic nature has not yet been fully characterised.}
   {We investigate the properties of the outer boundary of a long-lived sunspot, aiming to identify the magnetic property that defines it.}
   {We analysed the magnetic properties of AR NOAA\,11591 spot during its two passages across the solar disc. To this end, we used SDO/HMI continuum intensity and magnetic field parameters and determined the contours of these parameters to outline the outer boundary.}
   {During the first disc passage, in which the sunspot is in its stable phase, we find that the intensity contours at 0.9 of the mean intensity of the vicinity of the spot and isocontours of the magnetic field strength of 625\,G provide an almost perfect match between the two contours. With these thresholds, the time-averaged area of mismatch is minimised, yielding an average distance between the contours of 0.58 pixel, corresponding to less than 0.26 arcsec. During the second disc passage, the spot shows clear signs of decay, and we find that the 0.9 intensity and 625\,G magnetic isocontours detach from each other, coupled to the disappearance of penumbra. In this super-equipartition area, granulation still operates.}
   {Based on a comparison with simulation data from our previous work, and in agreement with findings of other authors, we conclude that the outer boundary of stable sunspots is defined by an invariant magnetic field: the equipartition field. From the discrepancy between intensity and magnetic contours during the decaying phase of the sunspot, we surmise that alongside the well-established (magneto-)convective regimes of the photosphere -- granular, penumbral, and umbral -- a super-equipartition granular regime can be identified. In this regime, bright, but smaller granules occur where the magnetic field exceeds equipartition but remains subcritical for convection suppression.}
   
   \keywords{  Sun: photosphere --
               Sun: magnetic fields --
               sunspots --
               magnetohydrodynamics (MHD) 
            }
   \authorrunning{Schmassmann, M., et al.}
   \titlerunning{A\&A, 706, A218 (2026)}
   \maketitle
   \hereOrThere{\myFooBar}{\nolinenumbers} 
%

\section{Introduction} 
\label{sec:intro}

Although sunspots are the largest photospheric manifestation of solar magnetism, the magnetic nature of the conspicuous boundary between two disparate modes of magnetoconvection, such as the umbra and penumbra of sunspots, has only recently been discovered and understood \citep{Jurcak:2011, Jurcak:2018, Schmassmann+2021}. This fundamental discovery was based on an analysis of the umbral boundary along its perimeter, defined by intensity contours \citep{Jurcak:2011}. Many previous attempts by other authors to characterise this boundary by fitting ellipses of increasing radius to investigate the radial dependence of magnetic properties failed to yield conclusive results. Following a similar approach as \citet{Jurcak:2011}, we studied the magnetic properties of the outer boundary of sunspots, that is, of sunspot penumbrae. 

Previously, from infrared observations of a symmetric sunspot, \citet{Solanki:1992} reported magnetic field strengths varying between 800\,G and 1000\,G along the sunspot outer boundary, depending on the location. Similar studies have followed with comparable results, analysing various sunspots and using different spectral lines \citep{Keppens:1996, cwp:2001, Bellot:2004}. With the first spectropolarimetric observations from satellites used to investigate the properties of the outer penumbral boundaries, the reported values of $B$ decreased to approximately 600\,G \citep{Borrero:2011, Jurcak:2011}. Yet, no conclusion about the nature of the outer boundary of sunspots was derived from these results.

\citet{Wiehr:1996} first pointed out a possible relation between a sharp intensity boundary of a sunspot and an equipartition field strength ($B_\mathrm{eq}$) that he estimated to be 750\,G, taking into account a density and velocity value from standard models of the solar photosphere. From low-resolution magnetograms and simultaneous photographic observations of a complex active region, \cite{Kalman:2002} found that the outer boundary of the sunspot penumbra closely follows the 750\,G isogauss line of the magnetic field strength and discussed its relation to the magnetopause of sunspot models by \cite{Jahn:1994} and the \mbox{relation} to the $B_\mathrm{eq}$ value. However, these two relevant results have received little attention in subsequent literature. 

\hereOrThere{\myTabTime}{\begin{table*}[ht]
	\caption{Timestamps of the studied spot during its two disc passages (year 2012).}
	\label{tab:time}
	\centering
	\begin{tabular}{c | c c c c | c @{} r r}
		\hline\hline\rule{0pt}{2.2ex}%
		NOAA AR & $t_\mathrm{<E}$ & $t_\mathrm{start}$ & 
		$t_\mathrm{end}$ & $t_{>W}$ & $t_\mathrm{ref}$ & 
		\rule{0pt}{2.2ex}Stonyhurst Lon & Lat\\
		\hline
        11591 & 10.11. 17:24 & 10.13. 19:48 & 10.22. 22:00 & 10.25. 08:00 & 10.17. 23:59:59Z & \rule{0pt}{2.2ex}$-7$ & +7\\
		11612 & 11.07. 22:00 &              &              & 11.21. 16:00 & 11.14. 23:59:59Z & +5 & +8\\
		\hline
	\end{tabular}\vspace*{1\baselineskip}
\end{table*}}

In a recent study, \cite{Jurcak+2020} compares the properties of various realistic MHD sunspot simulations and observations and shows (their Fig.~8, lower panels) that the sunspot magnetic field appears to be outlined by the equipartition field in the sub-photospheric layers and up to the surface. The authors also refer to this line, differentiating the sunspot field from the surrounding convective regime as the magnetopause. This finding further triggered the analysis presented here.
In magneto-hydrostatic sunspot models, such as the tripartite sunspot model of \cite{Jahn:1994}, the outer boundary outlined by the magnetopause is in total (gas plus magnetic) pressure equilibrium horizontally.

From a theoretical perspective, the equipartition magnetic field represents the characteristic field strength at which the magnetic energy density becomes comparable to the kinetic energy density of the plasma flow. This balance defines a physically meaningful threshold that separates the regimes dominated by fluid motion from those governed by magnetic forces. It can be derived that $B_\mathrm{eq}=\sqrt{4\pi\rho}\, v$, where $\rho$ is the density and $v$ is the \mbox{velocity}. The condition $B=B_\mathrm{eq}$ is equivalent to $v=$ $v_\mathrm{\scriptscriptstyle A}=B\big/\!\sqrt{4\pi\rho}$, whereby $v_\mathrm{\scriptscriptstyle A}$ is the {Alfv{\'e}n} speed. We note that the $B_\mathrm{eq}$ value cannot be accurately determined from observations, as only a few inversion codes provide reliable density values, and we cannot directly determine the velocity values, as the observed profiles are sensitive only to the line-of-sight component.

\section{Observations and data analysis}
\label{sect:obs_analys}
\myTabTime
\subsection{Data source and target} 
The investigation is based on the same dataset used by \cite{Schmassmann:2018}\defcitealias{Schmassmann:2018}{Paper~I}\citepalias[hereafter][]{Schmassmann:2018} for the study of the sunspot inner penumbral boundary. It consists of continuous observations with the Helioseismic and Magnetic Imager (HMI) on board NASA's Solar Dynamics Observatory (SDO) of a cutout of $500\times500$ pixels of a long-lived H-spot visible in NOAA AR 11591 and NOAA AR 11612 during its first and second disc passages, respectively, and with a time cadence of 12\,min. The sunspot shows a stable appearance throughout the first disc passage. During the second disc passage, the sunspot decays, as indicated by the fragmentation of its umbra into two parts near the central meridian and further decay later on \citepalias[see][\href{https://www.aanda.org/articles/aa/olm/2018/12/aa33441-18/aa33441-18.html}{movies corresponding to Fig.~3}]{Schmassmann:2018}

We applied the same methods used in \citetalias{Schmassmann:2018} to investigate the sunspot outer boundary. Here, we focused on analysing and comparing the contours of the intensity and the absolute magnetic field strength, $|B|$, at the outer spot boundary.

\subsection{Magnetic field maps}
The sequence of magnetic field maps was retrieved from the HMI inversions \citep{Borrero:2011a} and has been corrected for disambiguation and transformed into the local reference frame. Intensity maps have been corrected for centre-to-limb variation in the HMI pipeline, as explained in detail in \citetalias{Schmassmann:2018} and references therein.

\subsection{Identification of intensity contours}
The outer boundary of the sunspot observed with HMI can best be described at all times with an intensity contour that amounts to $I_\mathrm{c}=0.9\,I_\mathrm{qs}$, where $I_\mathrm{qs}$ is the average intensity of the surrounding granulation. 
The value of $0.9\,I_\mathrm{qs}$ was obtained from the minimisation of $\langle d\rangle_{\psi,t}$, which we introduce in Subsection~\ref{sect:quant_diff}.
Similarly to the methodology used in \citetalias{Schmassmann:2018}, to compensate for the different formation heights of the $I_\mathrm{c}$ and the \ion{Fe}{I} 617.3\,nm line, we transformed the coordinates obtained from the $I_\mathrm{c}=0.9\,I_\mathrm{qs}$ contours using
\begin{align}
	(x',y')&=\left(1+\frac{\Delta h}{R_\sun}\right)(x,y),\label{eq:trans}
\end{align}
before retrieving the magnetic field values at the boundary by linear interpolation, where $R_\sun=696\,$Mm, and $\Delta h$ is the height difference.
The inverse transform was applied when drawing magnetic field contours on the intensity maps. $\Delta h=255\,$km is another parameter that was obtained by minimising $\langle d\rangle_{\psi,t}$.

\subsection{Magnetic field time series fit}
\label{sect:time_series}
For each time step of the sunspot evolution, an average of the magnetic field along the contour was calculated, thus creating a time series of the form $X(t)=\left<|B|\right>_\psi(t)$ plotted with the results in Section~\ref{sec:B_temp} in Fig.~\ref{fig:1stH0.9}. Similarly, the standard deviations along the contours $\sigma_\psi(t)$ were calculated.
While we use the notation of averaging over the angle $\psi$, to differentiate it from the time direction, we actually averaged over the points along the contour, so the distances in $\psi$ between the points were not used as weights.
For the ranges $t_\mathrm{start}$ to $t_\mathrm{end}$ given in Table~\ref{tab:time}, this time series was then least-squares fitted using
\begin{align}
	X_\mathrm{fit}(t)&=X_0+X_3\cos(2\pi t-X_4),\label{eq:fit_t}
\end{align}
where $t$ is in days and $t\in\mathbb{N}$ is at noon. $X_0$ is the value we are interested in and is henceforth called the offset, and $X_3$ \& $X_4$ are the amplitude \& phase of the orbital artefacts introduced by the SDO's geosynchronous orbital motion \citepalias[see][]{Schmassmann:2018}. Furthermore, the root mean square deviation of the residuals 
\begin{align}{\sigma_t=\sqrt{\frac{1}{n}\sum_t 
\left( X(t)-X_\mathrm{fit}(t)\right)^2}}
\end{align}
was determined. $X_0$, $X_3$, $X_4$, $\sigma_t$, and the time average of the standard deviations along the contours $\langle\sigma_\psi\rangle_t$ are listed in Table~\ref{tab:res2}.
This offset $X_0$ from the first disc passage was then used as a contour level in the $|B|$ maps in Figs.~\ref{fig:scr1st0.9h_0}, \ref{fig:scr1st0.9h_rest}, and \ref{fig:scr2nd0.9h_all}, as well as the first paragraph of the following section.

\hereOrThere{\myFigFirstCentral}{\begin{figure}
	\includegraphics[width=\columnwidth]{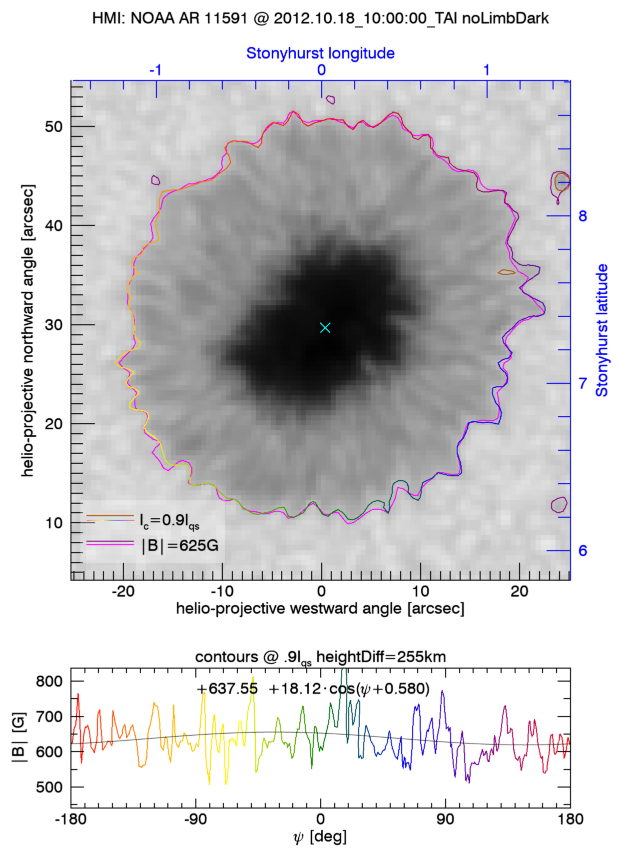} 
	\caption[continuum map and contours of NOAA AR 11591 at the central 
	meridian]
	{   NOAA AR 11591 at the central meridian. Top panel: 
		Continuum intensity map with a greyscale 
		$I_\mathrm{c}=0.1$ to $1.2\,I_\mathrm{qs}$. 
		The rainbow and brown contours outline $I_\textrm {c}=0.9\,I_\textrm{qs}$, where the colour of the rainbow marks the reference angle, $\psi$, with green pointing towards the disc centre. Light and dark purple contours outline $|B|=625\,$G.
		Different formation heights are accounted for 
		(Eq.~\eqref{eq:trans}, $\Delta h=255\,$km).
		The cyan cross denotes the umbral centroid.
		Lower panel: Absolute magnetic field \mbox{retrieved} along the 
		$I_\mathrm{c}=0.9\,I_\mathrm{qs}$ contour.\hspace*{3cm}\linebreak\null\hfill
        Video is available \href{https://www.aanda.org/10.1051/0004-6361/202558127/olm}{online}.
	}
	\label{fig:scr1st0.9h_0}
\end{figure}}

\hereOrThere{\myFigRadial}{\begin{figure}[t]
\includegraphics[width=1\linewidth]{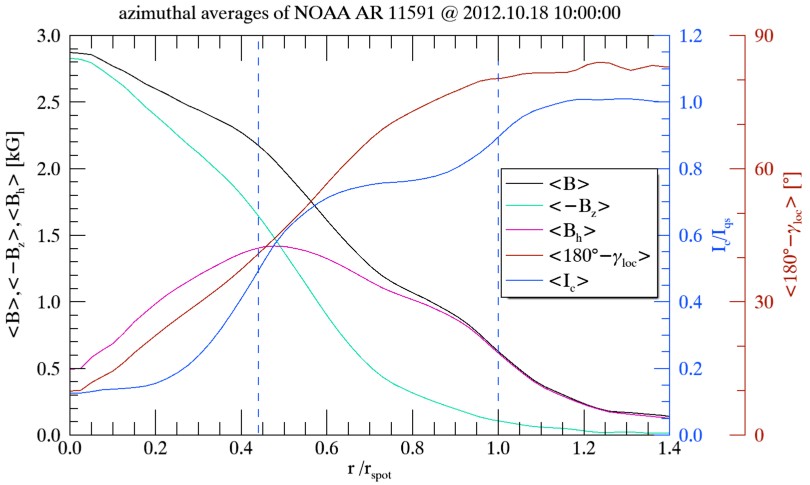}
\caption{Azimuthally averaged profiles of the sunspot magnetic field strength (black), vertical magnetic field (cyan), horizontal magnetic field (purple), field inclination (brown), and continuum intensity (blue) at the central meridian during the first disc passage. The vertical lines mark the umbral (0.5\,$I_\mathrm{qs}$) and spot (0.9\,$I_\mathrm{qs}$) boundaries. }
\label{fig:radialprofiles}
\end{figure}}

\hereOrThere{\myFigBabsFirst}{\begin{figure*}[ht]
	\sidecaption
	\includegraphics[width=12cm]{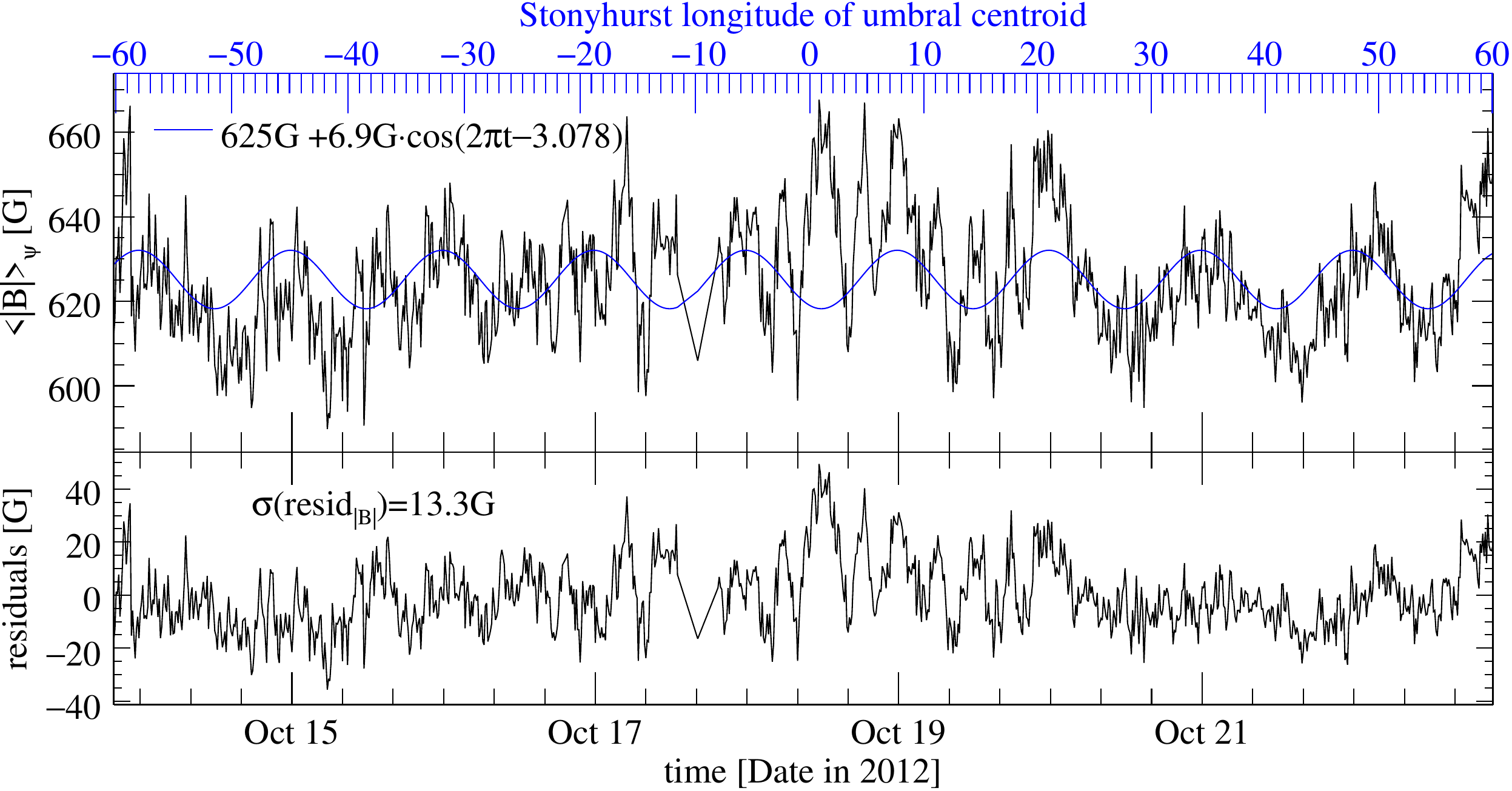}
	\caption[$\left<|B|\right>_\psi(t)$, where $I_\mathrm{c}=0.9\,I_\mathrm{qs}$ 
	and $\Delta h=255\,$km for NOAA AR 11591]%
    {Mean absolute magnetic field $\left<|B|\right>_\psi(t)$ along the $I_\mathrm{c}=0.9\,I_\mathrm{qs}$ contour, with $\Delta h=255\,$km accounted for, and the sinusoidal fits (top panel), and the residuals (bottom panel) for NOAA AR 11591 (1\textsuperscript{st} disc passage).}
    \label{fig:1stH0.9}
\end{figure*}}

\subsection{Quantifying the differences of intensity and magnetic contours}
\label{sect:quant_diff}
In Sect.~\ref{sec:results}, we compare two sets of contours: contours of intensity with contours at a fixed value of the field strength $|B|$. To quantify how well two sets of contours match, we define the average distance between two contours as $\langle d\rangle_{\psi}(t)={\Delta a(t)}\big/{\ell(t)}$, whereby $\Delta a(t)$ is the area of symmetric difference between two contours and $\ell(t)$ is the length of the intensity contour. 
When averaging in time, we used $\ell(t)$ as weights, resulting in $\langle d \rangle_{\psi,t}=\sum_t{\Delta a(t)}\big/\sum_t{\ell(t)}$, which are listed in Table~\ref{tab:res2}.

For every point along a contour, a reference angle, $\psi=\measuredangle(PCD)$, was calculated, whereby $P$ is the point on the contour, $C$ is the centroid of the $I_\mathrm{c}=0.5\,I_\mathrm{qs}$ contour in the intensity frame, and $D$ is the centre of the solar disc as observed by SDO. The angles were calculated on the sphere. For every time step the absolute magnetic field $Y(\psi)=|B|(\psi)$ at the  $I_\mathrm{c}=0.9\,I_\mathrm{qs}$ contour was least-squares fitted against functions of the form: 
\begin{align}
	Y_\mathrm{fit}(\psi)&=Y_0+Y_3\cos(\psi-Y_4)\label{eq:fit_p}.
\end{align}
The time average of the amplitude $\langle Y_3\rangle_t$ of these fits is listed in Table~\ref{tab:res2}, and the individual fits are plotted in the lower panel of Figs. ~\ref{fig:scr1st0.9h_0},~\ref{fig:scr2nd0.9h_all}~\&~\ref{fig:scr1st0.9h_rest} and the \href{https://www.aanda.org/10.1051/0004-6361/202558127/olm}{videos}.

\subsection{Optimisation for the difference in formation height} 
\label{sect:diff_height}
We determined the optimal height difference $\Delta h=255\,$km used in Eq.~\eqref{eq:trans} by minimising the average distance, $\langle d\rangle_{\psi,t}$, between intensity contours and contours at a fixed value of $|B|$. We varied both $\Delta h$ and the intensity threshold simultaneously during the minimisations (see Table \ref{tab:res2}). As described above, the threshold for the $|B|$ contour is a function of the intensity threshold and $\Delta h$.
In \citetalias{Schmassmann:2018}, the value of $\Delta h$ considered was higher to account for the Wilson depression and differential line-of-sight effects.
Here, only the projection effects due to the difference in formation height of the continuum and the \ion{Fe}{I} 617.3\,nm line are relevant. This $\Delta h$ is consistent with expectations from radiative transfer calculations \citep[see e.g.][Table 1]{Norton+2006}.
In the time range we evaluated, the heliocentric angle of the spot is below $60\degr$, and the impact of changing $\Delta h$ on $\langle d\rangle_{\psi,t}$ is small, but closer to the limb, it becomes very pronounced. This is expected, as $\Delta h=255\,$km corresponds to 0.7 pixels there.

By inspecting the contours with $\Delta h$ set to $0$\,km, and hence, with no compensation for projection effects, we observe that with increasing heliocentric angles, the magnetic contour is \mbox{increasingly} shifted away from disc centre relative to the intensity contour. Therefore, on the disc (limb) side of the spot, the intensity contour is outside (inside) the magnetic contour, and the magnetic field at the location of the intensity contour is reduced (increased). This results in a large sinusoidal variation of $Y(\psi)=|B|(\psi)$ retrieved along the intensity contour, with a phase $Y_4\approx\pm\pi$. The amplitude $Y_3$ increases with heliocentric angle, and this is reflected in larger values of $\langle Y_3\rangle_t$ for $\Delta h=0$\,km relative to $\Delta h=255$\,km in Table \ref{tab:res2}.
It is reassuring to note that neglecting the projection effect does neither influence the mean value around the contour nor $Y_0$, as the limb-side and disc-side sinusoidal variation cancels out.

\section{Results}
\label{sec:results}
\myFigFirstCentral
Figure~\ref{fig:scr1st0.9h_0} summarises the main result of this investigation. For the stable spot near disc centre, the match between the isocontour at the $I_\mathrm{c}=0.9\,I_\mathrm{qs}$ intensity threshold outlining the spot outer boundary and the magnetic isocontour at $|B|=625$\,G is remarkable, considering that the threshold value of 625\,G was obtained by averaging in time for more than 9 days. The corresponding procedure to obtain the best intensity threshold and the resulting value for the magnetic field strength is described in Section~\ref{sect:quant_diff}. For this particular time step, the average of $B$ along the contour is 638\,G. This deviates from the time-averaged value of 625\,G due to orbital artefacts, as mentioned in Sect.~\ref{sect:time_series}.
In the following, we present in more detail this and other results (on the vertical component of the magnetic field) following the evolution of this long-lived sunspot across its first (stable phase) and second \mbox{(decay)} disc passages.

\subsection{Radial distribution of the magnetic parameters}
\myFigRadial
Figure~\ref{fig:radialprofiles} shows the radial profiles of the absolute magnetic field strength ($|B|$), vertical ($B_\mathrm{z}$) and horizontal ($B_\mathrm{h}$) components, field inclination ($180\degr - \gamma_\mathrm{\scriptscriptstyle LRF}$), and continuum intensity ($I_\mathrm{c}$) of the sunspot when located at the central meridian. For this purpose, we calculated the azimuthal averages of the magnetic parameters for circular cuts around the spot centroid \mbox{extending} up to 1.4 times the spot radius. The plot confirms the well-established global properties of sunspots \citep[see e.g. the reviews by][]{Borrero:2011, Solanki:2003}: intensity increases with radius from umbral values through the sharp umbra-penumbra boundary around 0.5\,$I_\mathrm{qs}$, to the intensity plateau through the penumbral area around 0.75\,$I_\mathrm{qs}$, to the less conspicuous yet apparent outer penumbral boundary around 0.9\,$I_\mathrm{qs}$. 
The absolute magnetic field strength drops from the umbral (maximum) values around 2.8\,kG--2.5\,kG at the umbral boundary, and to around 600\,G at 0.9\,$I_\mathrm{qs}$. 
In contrast, the vertical field drops more rapidly from 2.7\,kG in the centre of the umbra to around 1.7\,kG at the umbral boundary -- corresponding to the Jur\v cák canonical value \citepalias[see][]{Schmassmann:2018} -- and down to around 100\,G at 0.9\,$I_\mathrm{qs}$. 
The field inclination varies according to $B_\mathrm{z}$: highly vertical fields (<20$\degr$, with respect to the normal to the solar surface) at the central umbra, about 45$\degr$ at the umbral boundary, and highly horizontal (80$\degr$) at the sunspot boundary. 
Redundantly, $B_\mathrm{h}$ increases with the sunspot radius to a maximum of 1.4\,kG at the umbral boundary, then decreases along the penumbra until reaching $|B|$ values at 0.9\,$I_\mathrm{qs}$  of the spot radius and beyond.

However, the analysis of the magnetic field parameters with the spot radius has historically eluded an explanation of the magnetic nature of the umbral and sunspot boundaries. Therefore, in the following, we focus this analysis on the investigation of the magnetic properties along the boundary.

\hereOrThere{\myFigSecond}{\begin{figure*}
\centering
\includegraphics[trim= -0.025in 0pt -0.025in 0pt, clip, width=0.49\textwidth]{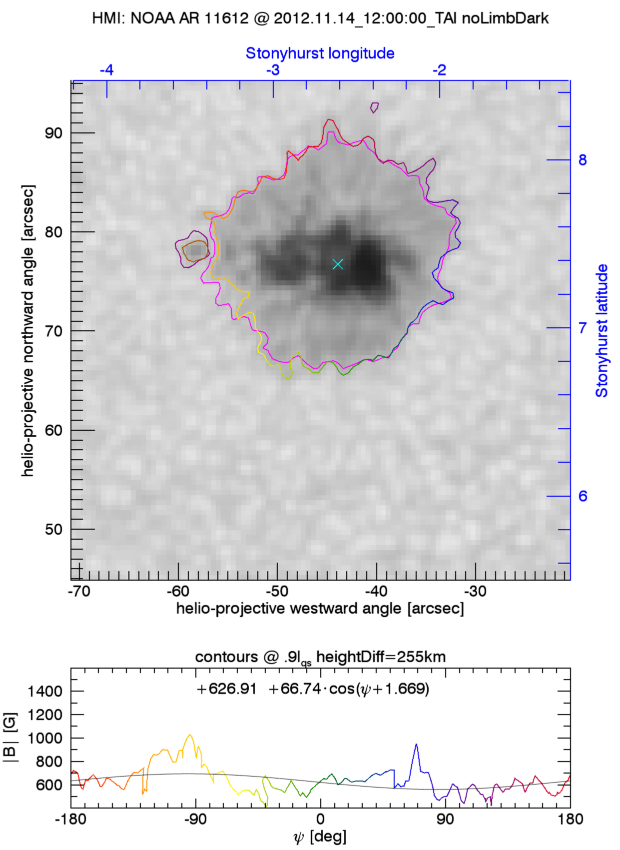}%
\includegraphics[trim= -0.025in 0pt -0.025in 0pt, clip, width=0.49\textwidth]{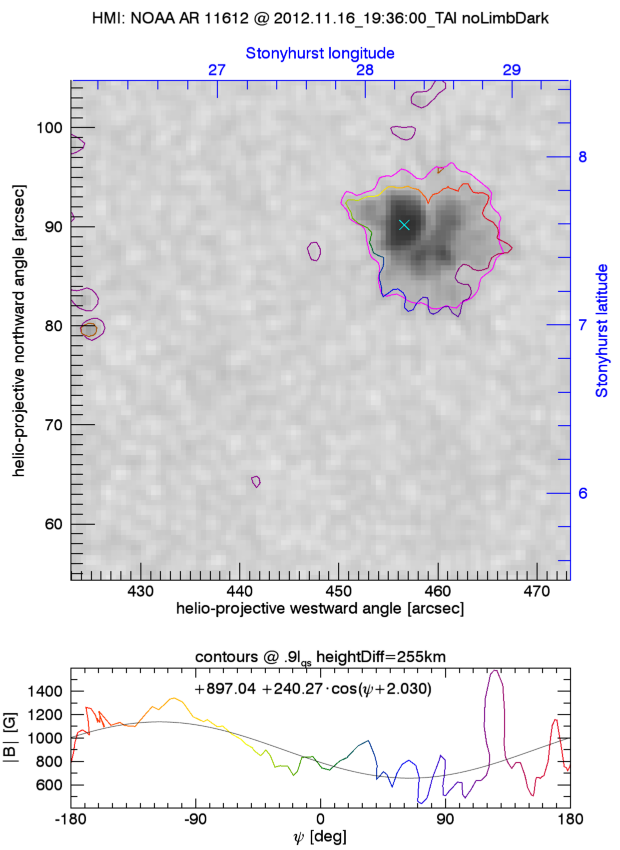}%
\caption{Same as Fig.~\ref{fig:scr1st0.9h_0}, but for NOAA AR 11612 (2\textsuperscript{nd} disc passage) at Stonyhurst longitudes of the centroid near $-3\degr$ (left) and $28\degr$ (right).\hphantom{Video}\linebreak
\null\hfill\mbox{Video is available \href{https://www.aanda.org/10.1051/0004-6361/202558127/olm}{online}.}}
\label{fig:scr2nd0.9h_all}
\end{figure*}}

\hereOrThere{\myFigFieldIncl}{\begin{figure*}
\centering
\includegraphics[width=0.49\textwidth]{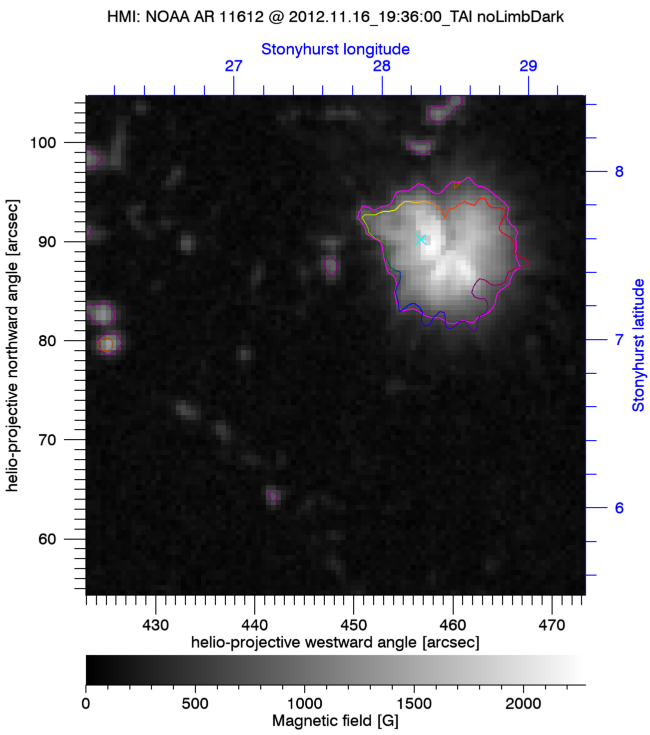}%
\includegraphics[width=0.49\textwidth]{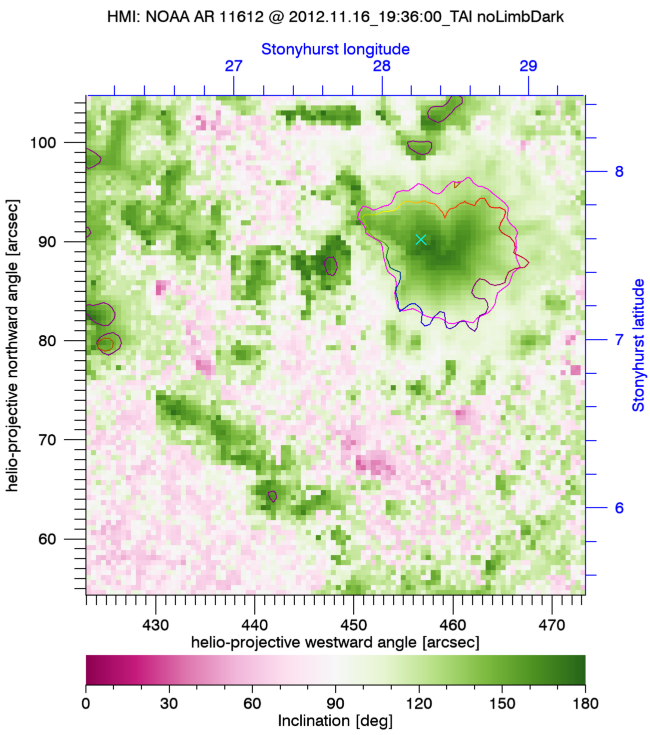}%
\caption{Same time as the right panels of Fig.~\ref{fig:scr2nd0.9h_all}, but showing magnetic field strength, $|B|$ (left), and inclination, $\gamma_\mathrm{\scriptscriptstyle LRF}$ (right).}
\label{fig:scr2nd0.9h_mag_incl}
\end{figure*}}

\subsection{Temporal evolution of |B| along the outer sunspot boundary.} 
\label{sec:B_temp}
\myFigBabsFirst
Figure~\ref{fig:1stH0.9} shows the azimuthally averaged $|B|$ along the 0.9\,$I_\mathrm{qs}$ intensity contour over the course of the first disc passage, applying the fit described in Sect.~\ref{sect:time_series}. The offset parameter $X_0$, corresponding in this case to $|B|$, is 625\,G. The plot includes the correction for the difference in formation height ($\Delta h = 255$\,km; see Sect.~\ref{sect:diff_height}). The lower panel displays the residuals after subtracting the fitted curve. 
For completeness, Table~\ref{tab:res2} in the Appendix lists the parameters of the sinusoidal fits along with the rms values of the residuals for $|B|$, $\sigma_t$, calculated using various intensity thresholds around 0.9\,$I_\mathrm{qs}$.
For fixed $\Delta h=255\,$km, 0.9\,$I_\mathrm{qs}$ is the threshold value resulting in the lowest $\sigma_t=13.3\,$G, which validates our choice of threshold.
Note that while this indicates that the outer spot boundary is on average very stable, there are fluctuations on the scale of individual convective cells, resulting in an average rms along the contour of $\langle\sigma_\psi\rangle_t=73.6\,$G.

\fontdimen4\font=1pt\looseness=-1%
For comparison, we also provide the fit and rms values for the vertical magnetic field component, $B_\mathrm{ver}$, and the inclination in the local reference frame, $\gamma_\mathrm{\scriptscriptstyle LRF}$, in Table~\ref{tab:res5}. Here, we used a slightly shortened time range to avoid erroneous disambiguation results, where the heliocentric angle approaches $60\degr$. 
We see that the vari\-ation of $B_\mathrm{ver}$ along the contour is almost as large as its average. The field is on average only $12\degr$ above horizontal, and the varia\-tion along the contour is not much less than that. Hence, neither $B_\mathrm{ver}$ nor the field inclination is suitable for characterising the outer%
\linebreak\fontdimen2\font=1pt%
boundary of a sunspot. Because the field is close to horizontal, $B_\mathrm{hor}$ is very similar to $|B|$. While we cannot exclude that $B_\mathrm{hor}$ would be%
\linebreak\fontdimen2\font=1.5pt%
equally suited as $|B|$ to delineate the outer boundary of a spot, we see no theoretical reasons to expect so, and therefore see no reason%
\linebreak\fontdimen4\font=0.59999pt\fontdimen2\font=2.5pt
to use values sensitive to faulty $180\degr$-disambiguation.

\subsection{Contours} 
\label{sect:contours}
We show contours corresponding to both the 0.9\,$I_\mathrm{qs}$ intensity level and the $|B|=625$\,G magnetic field strength for the first disc passage in Fig.~\ref{fig:scr1st0.9h_0} and the related \href{https://www.aanda.org/10.1051/0004-6361/202558127/olm}{online movie}. Selected snapshots are also available in the Appendix (Fig.~\ref{fig:scr1st0.9h_rest}, first four panels). For the second disc passage, during which the sunspot decays, Fig.~\ref{fig:scr2nd0.9h_all} shows the selected snapshots discussed below. A movie of the full disc passage is also available \href{https://www.aanda.org/10.1051/0004-6361/202558127/olm}{online} and \mbox{selected} snapshots are available in Fig.~\ref{fig:scr1st0.9h_rest} (last two panels).

\paragraph{Sunspot stable phase -- first disc passage.} During the first disc passage, the sunspot shows clear signs of stability. During this period, the $|B|=625$\,G isocontour consistently and accurately traces the sunspot's outer boundary, confirming its reliability as a stable sunspot boundary marker for SDO/HMI data. 
While this can be clearly seen in the video, it is also quantitatively confirmed by calculating the average distance between contours, $\langle d\rangle_{\psi,t}$, which shows that on average the $I_\mathrm{c}=0.9\,I_\mathrm{qs}$ contours and the $|B|=625$\,G contours are only 0.575 pixels apart, which corresponds to less than 0.26 arcsec. We note that at the limb, the projection effects from the formation heights are more significant.

\myFigSecond
\myFigFieldIncl
\paragraph{Sunspot decay phase -- second disc passage.} During the second disc passage, however, Fig.~\ref{fig:scr2nd0.9h_all} (and the related \href{https://www.aanda.org/10.1051/0004-6361/202558127/olm}{online movie}) shows that both the 0.9$\,I_\mathrm{qs}$ and $|B|=625$\,G contours \mbox{begin} to detach from the spot boundary, coinciding with clear signs of sunspot decay. This is also reflected in the evolution of the average $|B|$ values displayed in Fig.~\ref{fig:2nd_passage}, available in the \mbox{Appendix}. This occurs during the disappearance of the penumbra, that is, owing to the disappearance of the horizontal field, possibly turning more vertical \citep[see e.g.][]{Bellot+2008}, during the spot decay. At that stage, the $0.9\,I_\mathrm{qs}$ retracts closer to the umbra, but a 625\,G isocontour is still present some 2\,arcsec further out in the naked (without penumbra) area (see top right panel of Fig.~\ref{fig:scr2nd0.9h_all}). Inside that area, between the 0.9$\,I_\mathrm{qs}$ and $|B|=625$\,G contours, the field remains below $1600\,$G (see Fig.~\ref{fig:scr2nd0.9h_all} bottom right panel for $|B|$ along the intensity contour, and the magnetic field map in Fig.~\ref{fig:scr2nd0.9h_mag_incl}, left panel), and is less inclined than in the penumbra (see Fig.~\ref{fig:scr2nd0.9h_mag_incl}, right panel). 
While the field there is too weak to suppress convection as in the umbra in the context of the Jurčák criterion \citepalias[see][]{Schmassmann:2018}, it is super-equipartition, but nevertheless, it has apparently no significant effect on convection.
This means that granulation fills the gap between the new spot boundary and the 625\,G isocontour, which shows no corresponding change in the intensity.

\section{Discussion and conclusion} 
We find that for a stable sunspot, a canonical value of the magnetic field strength defines its outer intensity edge. This canonical value corresponds to a well-defined physical quantity. \citet{Wiehr:1996} and later \citet{Kalman:2002} have already pointed out that the outer boundary of sunspots is given by the equipartition field.
Indeed, from an analysis of sunspot simulations, \cite{Jurcak+2020} finds that the edge of the sub-photospheric sunspot magnetic trunk, also referred to in magneto-hydrostatic models as the magnetopause, is defined by the equipartition field at all depths and up to the `solar' surface, delimiting the sunspot boundary (see their Figs.~6 and 8). The equipartition field is quantified in their potential sunspot simulation to be 750\,G (at $\tau =1$), similar to that estimated in convective simulations \cite{Rempel2014} and that of standard models \citep{Wiehr:1996}.
The equipartition field defines the critical value at which the kinetic energy density and magnetic energy density are balanced. This means that in super-equipartition areas (with $|B| > B_\mathrm{eq}$), the magnetic field dominates over convection, whereas in sub-equipartition areas (with $|B| < B_\mathrm{eq}$), convection governs over magnetic fields.  

Since, in sunspots, the azimuthally averaged vertical field \mbox{decreases} rapidly with increasing radius, the values of $|B|$ and 
$B_\mathrm{hor}$ becomes nearly identical at the outer boundary (Fig.~\ref{fig:radialprofiles}). However, and according to the theory, we argue that it is the total field strength, $|B|$, reaching the equipartition field that defines the visible edge of the spot.

It is important to note that convective cells form deep within the convection zone, whereas the magnetic field is observed in the photosphere. Consequently, while the condition $|B|=B_\mathrm{eq}$ may define the spot boundary below the atmosphere, the \mbox{observed} $|B|$ depends on the height at which the measurement is sensitive. Therefore, observations at different wavelengths --\linebreak as well as other aspects of the observing and data-processing methods -- will affect the $|B|$ value that delineates the spot boundary.
Thus, we claim the invariance of the magnetic value at the sunspot boundary, yet the estimated value of $|B|=625$\,G is specific to the data and analysis used in this study and shall vary in other studies \citep[e.g.][]{Kalman:2002}.

\setlength{\baselineskip}{11pt plus 1pt}
\paragraph{The equipartition field in sunspots.} According to our results and a previous study on simulations \citep{Jurcak+2020}, the equipartition field in sunspots defines the outer boundary of the penumbra, indicating that within this region, the magnetic field dominates over convection: The penumbra represents the part of a sunspot where the vertical magnetic field is weak enough to permit overturning convection \citep{jurcak:2014, Jurcak:2015, Jurcak:2017, Jurcak:2018, Schmassmann+2021}, while concurrently strong horizontal fields shape this convection into a filamentary form of magnetoconvection up to a radial distance at which the equipartition limit is reached. Beyond the sunspot boundary, the sub-equipartition field is governed by the surrounding granulation. 

The fact that the outer boundary of the penumbra is \mbox{defined} by a canonical value -- the equipartition field -- may \mbox{appear} trivial within the common framework of sunspot models. However, this is highly significant, as it outlines the surface intersection of the sub-photospheric magnetic trunk of a super-equipartition magnetic structure.
In the case of sunspots, this condition effectively predetermines the average radial extent (or length) of their penumbrae. Observational studies consistently report that penumbral extensions amount to approximately 0.6 times the spot radius, independent of the overall sunspot size \citep[e.g.][]{Keppens:1996,cwp:2001, Mathew:2003, Borrero:2004, Bellot:2004, Sanchez:2005, Beck:2008, Borrero:2011}.
Because the length of individual penumbral filaments is known to be limited to a maximum of about 6500\,km \citep{Tiwari:2013}, an additional mechanism is required to maintain the 0.6 ratio in large sunspots. Indeed, larger sunspots, which host stronger magnetic fields, exhibit broader penumbrae composed of two or more radially interlaced rows of filaments, which extend outwards until the equipartition boundary of the spot is reached. 

\paragraph{Extrapolation to the case of pores and other super-equipartition photospheric structures.}
In pores, the typically reduced magnetic flux (compared to sunspots) leads to a more vertical magnetic-field topology and, consequently, a distinct morphological appearance \citep[see e.g.][]{Keppens:1996}. The lack of a strong horizontal component results in an annular region, rather than a penumbra, surrounding the visible pore, where the predominantly vertical magnetic fields exceed 1 kG \citep[see e.g.][]{Keppens:1996, GarciaRivas+2024, Verma2024}. In these regions, super-equipartition magnetic fields dominate the local dynamics. How\-ever, these fields, although stronger than equipartition, \mbox{remain} below the Jurčák critical value \citepalias[see][]{Schmassmann:2018} and therefore do not suppress convection as in the umbra. As a result, granulation continues to operate around pores, resembling the `naked' areas of decaying sunspots described in Sect.~\ref{sect:contours}. The influence of such super-equipartition fields typically manifests as granulation with reduced cell size as found around sunspots, in granular light bridges, and in plage regions \citep{Macris:1979, Hirzberger:2002, Narayan+Scharmer2010}.
In this context, alongside the well-established (magneto-)convective regimes of the photosphere -- granular, penumbral, and umbral 
\linebreak\fontdimen2\font=3.5pt\fontdimen3\font=3pt%
-- a super-equipartition granular regime can be identified. In this regime, bright but smaller granules occur where the mag\-netic field exceeds equipartition, but remains subcritical to suppress overturning convection, and is unable to elongate the convective cells, possibly due to a too weak horizontal 
\fontdimen2\font=2.5pt\fontdimen3\font=1.49998pt%
component.

\setlength{\baselineskip}{11pt}
\paragraph{Conclusion and outlook.}  We present a case study investigating an invariant magnetic field strength that defines the outer boundary of sunspots. Our results and those of our previous studies suggest that this invariant field corresponds to the equipartition field. Verifying these findings and testing this \mbox{conjecture} requires two key follow-up analyses.
First, a comprehensive statistical study of sunspot outer boundaries across different evolutionary stages -- from pores to mature sunspots and, \mbox{ultimately}, to naked spots -- is required. 
Second, although direct \mbox{observational} measurements of the equipartition field are not yet available, inversion codes enable the retrieval of plasma densities. With the continued development of inversion tools such as FIRTEZ \citep{PastorYabar+2019, Borrero+2019}, currently being advanced by \citet{VillaCrespoSubmitted}, it will also become possible to estimate horizontal velocities from observations, thereby allowing an empirical quantification of the equipartition field.

\section*{Data availability}
Movies associated to Figs.~\ref{fig:scr1st0.9h_0} and \ref{fig:scr2nd0.9h_all} are available on \href{https://arxiv.org/abs/2512.11160}{arXiv} with the source files and at \\
\resizebox{\linewidth}{!}{\url{https://www.aanda.org/10.1051/0004-6361/202558127/olm}}

\begin{acknowledgements}
We thank the anonymous referee for their helpful suggestions.
This work was supported by the Czech-German common grant, funded by the Czech Science Foundation under the project 23-07633K and by the Deutsche Forschungsgemeinschaft under the project BE 5771/3-1 (eBer-23-13412), and the institutional support ASU:67985815 of the Czech Academy of Sciences.
This research has made use of NASA's Astrophysics Data System Bibliographic Services.
\end{acknowledgements}

\noindent\textcolor{red}{This version of the article differs from the published A\&A version only by typesetting.}\\
\newpage
\bibliographystyle{aa_url}
\begin{hyphenrules}{nohyphenation}
\bibliography{biblio}
\end{hyphenrules}

\begin{appendix}
\onecolumn
\section{Tables of fit parameters}

\begin{table*}[htp!]
\centering
\caption{Fit parameters for $\left<|B|\right>_\psi(t)$ [G] 
	and time averages across the first disc passage}
\label{tab:res2}
\hfill\begin{math}\begin{array}{@{}r r || r r r r | r | r | r | r r || r }
	\hline\hline
I_\mathrm{c}/I_\mathrm{qs} & \Delta h[\text{km}] & X_0 & X_3 & X_4\iflatexml\textsuperscript{\textcolor{blue}{a}}\else\fi\tablefootmark{a}
& \sigma_t & \langle\sigma_\psi\rangle_t & \left<Y_3\right>_t & 
\left\langle d\right\rangle_{\psi,t}\iflatexml\textsuperscript{\textcolor{blue}{b}}\else\fi\tablefootmark{b}\hspace*{-3.1pt} & \sigma_t/X_0 & 
\langle\sigma_\psi\rangle_t/X_0 & 
\langle r_\mathrm{spot}\rangle_t\iflatexml\textsuperscript{\textcolor{blue}{c}}\else\fi\tablefootmark{c}
	\rule[-1ex]{0pt}{3.2ex}\\
	\hline
	\rule{0pt}{2.2ex}
 0.87 &   0 & 684.6 & 5.4 & -3.114 & 15.7 & 81.5 & 43.1 & 0.6307 & 0.02286 & 0.11912 & 14.29\\
 0.88 &   0 & 664.6 & 5.4 & -3.129 & 14.9 & 80.4 & 43.2 & 0.6220 & 0.02245 & 0.12093 & 14.40\\
 0.89 &   0 & 645.0 & 5.7 &  3.090 & 14.2 & 80.2 & 44.0 & 0.6168 & 0.02203 & 0.12434 & 14.50\\
 {\bf 0.90} &   {\bf 0} & {\bf 625.3} & {\bf 6.0} &  {\bf 3.073} & {\bf 13.7} & {\bf 80.3} & {\bf 44.7} & {\bf 0.6161} & {\bf 0.02193} & {\bf 0.12838} & {\bf 14.61}\\
 0.91 &   0 & 605.3 & 6.3 &  3.055 & 13.3 & 80.7 & 45.4 & 0.6198 & 0.02199 & 0.13337 & 14.72\\
 0.92 &   0 & 585.2 & 6.5 &  3.015 & 13.0 & 81.8 & 46.5 & 0.6261 & 0.02227 & 0.13979 & 14.83\\
 0.93 &   0 & 564.1 & 6.8 &  3.008 & 12.7 & 82.7 & 47.4 & 0.6381 & 0.02254 & 0.14660 & 14.95\\
 \hline
 \rule{0pt}{2.2ex}
 0.87 & 255 & 683.8 & 6.2 & -3.110 & 15.3 & 75.9 & 24.1 & 0.5984 & 0.02243 & 0.11104 & 14.29\\
 0.88 & 255 & 664.1 & 6.3 & -3.132 & 14.5 & 74.4 & 23.5 & 0.5867 & 0.02184 & 0.11200 & 14.40\\
 0.89 & 255 & 644.7 & 6.6 &  3.095 & 13.8 & 73.9 & 23.2 & 0.5787 & 0.02137 & 0.11456 & 14.50\\
 {\bf 0.90} & {\bf 255} & {\bf 625.1} & {\bf 6.9} &  {\bf 3.078} & {\bf 13.3} & {\bf 73.6} & {\bf 23.2} & {\bf 0.5751} & {\bf 0.02122} & {\bf 0.11776} & {\bf 14.61}\\
 0.91 & 255 & 605.3 & 7.1 &  3.053 & 12.9 & 73.8 & 23.2 & 0.5757 & 0.02133 & 0.12191 & 14.72\\
 0.92 & 255 & 585.4 & 7.3 &  3.020 & 12.7 & 74.6 & 23.6 & 0.5791 & 0.02170 & 0.12745 & 14.83\\
 0.93 & 255 & 564.4 & 7.6 &  3.003 & 12.5 & 75.3 & 24.0 & 0.5884 & 0.02220 & 0.13349 & 14.95\\
 \hline
\end{array}\end{math}\hfill\null\newline
\tablefoot{
\tablefoottext{a}{in radian, $X_4\approx\pm\pi$ means 
	$X_\mathrm{fit}(t)$ is largest when ${t\approx0.5+\mathbb{N}}$, 
	i.e. around midnight.}\\
\tablefoottext{b}{average distance between two contours in pixel, see text}\quad
\tablefoottext{c}{in Mm, $r_\mathrm{spot}=\sqrt{a_\mathrm{deprojected}/\pi}$}}
\end{table*}

\begin{table*}[htp]
	\caption{Fit parameters for $\left<-B_\mathrm{ver}\right>_\psi(t)$ \& 
		$\left<\gamma_\mathrm{\scriptscriptstyle LRF}\right>_\psi(t)$ [G] and time averages
		(only 10.14 23:36 to 10.21 19:00 used)}
	\label{tab:res5}
	\centering
	\hfill\begin{math}\begin{array}{r r || r r r r | r | r || r r r r | r | r }
		\hline\hline
    \multicolumn{2}{c||}{\rule[-1ex]{0pt}{3ex}} &
    \multicolumn{6}{c||}{\left<-B_\mathrm{ver}\right>_\psi(t)\text{ [G]}} &
    \multicolumn{6}{c}{\left<\gamma_\mathrm{\scriptscriptstyle LRF}\right>_\psi(t)\ [\degr]}\\
    \hline
		I_\mathrm{c}/I_\mathrm{qs} & \Delta h[\text{km}] & X_0 & X_3 & X_4\iflatexml\textsuperscript{\textcolor{blue}{a}}\else\fi\tablefootmark{a} 
		& \sigma_t & \langle\sigma_\psi\rangle_t & \left<Y_3\right>_t & 
		X_0 & X_3 & X_4\iflatexml\textsuperscript{\textcolor{blue}{a}}\else\fi\tablefootmark{a}
		& \sigma_t & \langle\sigma_\psi\rangle_t & \left<Y_3\right>_t
        \rule[-1ex]{0pt}{3.2ex}\\
		\hline
		\rule{0pt}{2.2ex}
 0.87 & 255 & 149.5 & 3.5 &  2.994 & 13.9 & 84.9 & 57.4 & 102.5 & 0.27 &  3.023 & 1.06 & 6.60 & 4.87 \\
 0.88 & 255 & 143.3 & 3.7 &  3.043 & 13.5 & 83.3 & 56.4 & 102.4 & 0.28 &  3.082 & 1.08 & 6.70 & 4.95 \\
 0.89 & 255 & 137.5 & 3.6 &  2.982 & 12.9 & 82.2 & 55.9 & 102.2 & 0.26 &  3.044 & 1.08 & 6.85 & 5.05 \\
 {\bf 0.90} & {\bf 255} & {\bf 131.5} & {\bf 3.8} &  {\bf 3.034} & {\bf 12.4} & {\bf 80.7} & {\bf 55.6} & {\bf 102.1} & {\bf 0.27} &  {\bf 3.113} & {\bf 1.08} & {\bf 6.97} & {\bf 5.15} \\
 0.91 & 255 & 125.7 & 4.0 &  2.999 & 12.3 & 79.9 & 55.3 & 101.9 & 0.28 &  3.103 & 1.12 & 7.14 & 5.27 \\
 0.92 & 255 & 119.9 & 3.6 &  3.028 & 12.5 & 79.5 & 55.1 & 101.8 & 0.25 & -3.123 & 1.16 & 7.34 & 5.40 \\
 0.93 & 255 & 113.8 & 3.8 &  3.061 & 12.0 & 78.6 & 54.9 & 101.6 & 0.26 & -3.073 & 1.17 & 7.55 & 5.54 \\
 \hline
	\end{array}\end{math}\hfill\null\newline
\tablefoot{
\tablefoottext{a}{in radian, $X_4\approx\pm\pi$ means 
	$X_\mathrm{fit}(t)$ is largest when ${t\approx0.5+\mathbb{N}}$, 
	i.e. around midnight.}}
\end{table*}

\section{Mean magnetic field versus time during the \texorpdfstring{2\textsuperscript{nd}}{2nd} disc passage}

\begin{figure*}[htp]
	\sidecaption
	\includegraphics[width=12cm]{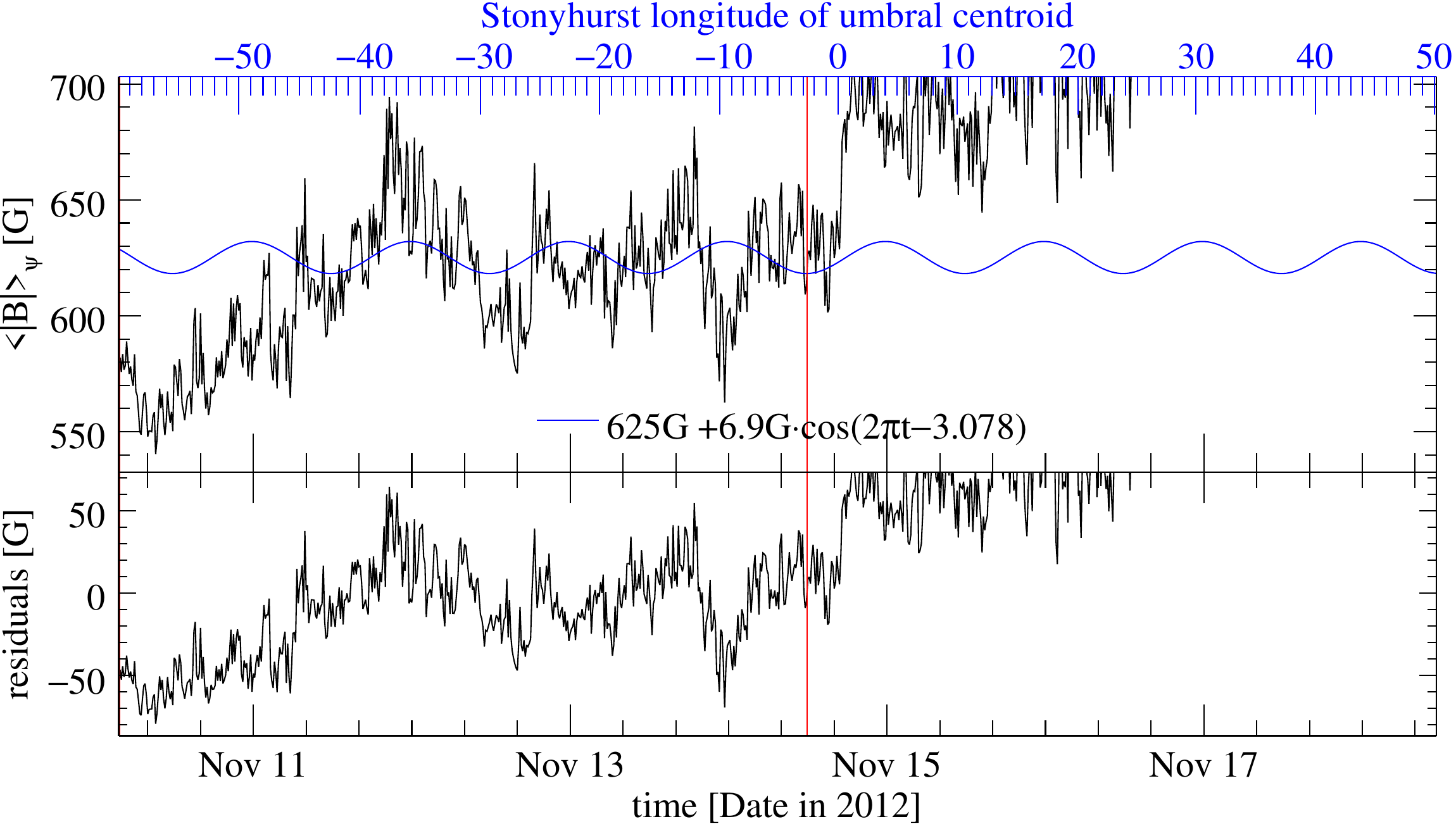}
	\caption[$\left<|B|\right>_\psi(t)$, where $I_\mathrm{c}=0.9\,I_\mathrm{qs}$ 
	and $\Delta h=255\,$km for NOAA AR 11612]%
    {Same as Fig.~\ref{fig:1stH0.9}, but for the 2\textsuperscript{nd} disc passage (NOAA AR 11612).
    Mean absolute magnetic field $\left<|B|\right>_\psi(t)$ along the $I_\mathrm{c}=0.9\,I_\mathrm{qs}$ contour, with $\Delta h=255\,$km accounted for, and the sinusoidal fits from the 1\textsuperscript{st} disc passage (top panel), and the residuals (bottom panel). 
    The red line indicates the time when the umbra falls apart.}
	\label{fig:2nd_passage}
\end{figure*}

\newpage
\section{Figures of isocontours for selected heliocentric angles}

\begin{figure*}[htp]
\centering
\includegraphics[width=0.33\textwidth]{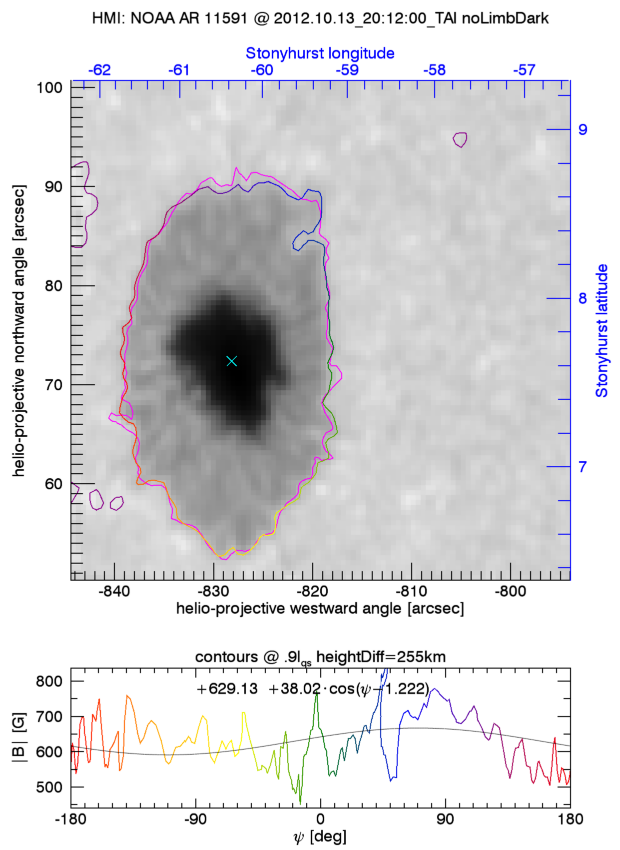}%
\includegraphics[width=0.33\textwidth]{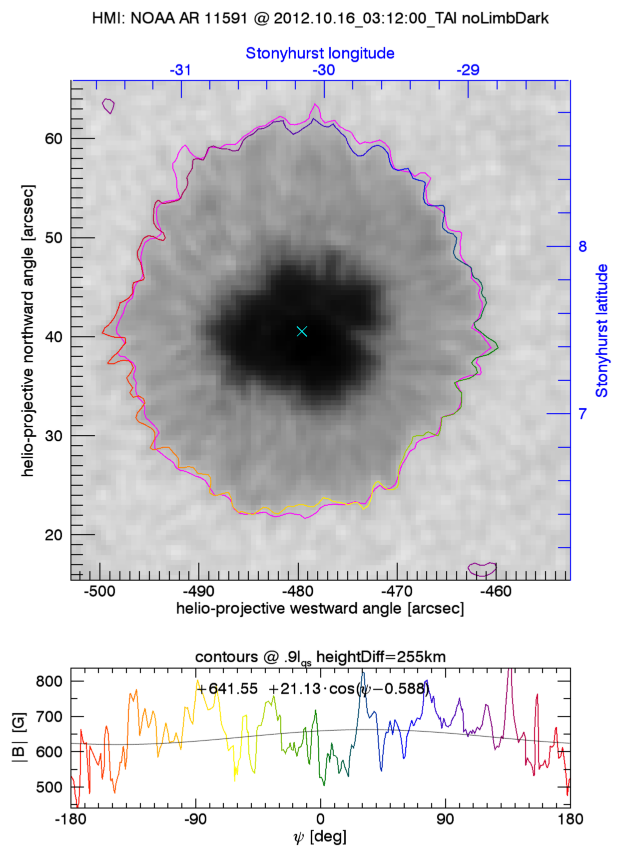}%
\includegraphics[width=0.33\textwidth]{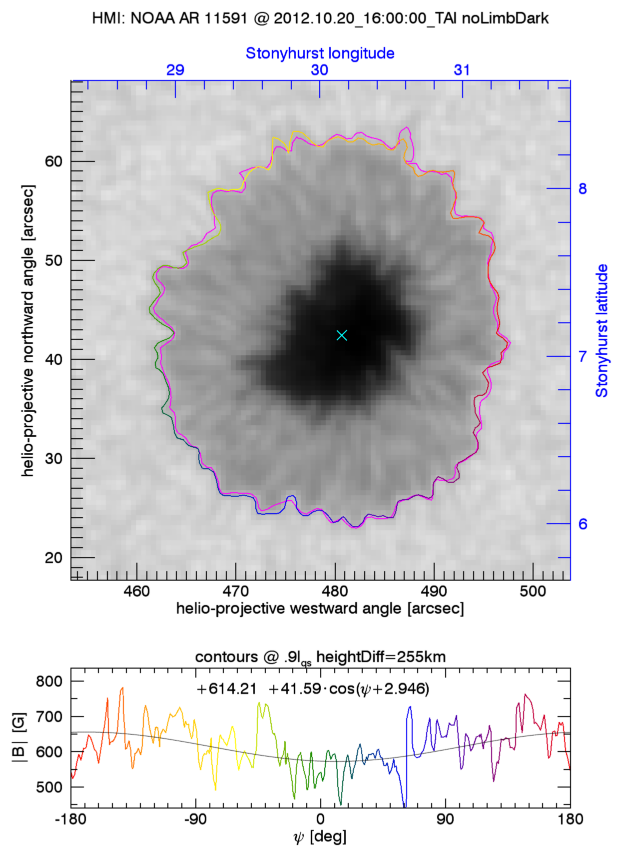}\\
\includegraphics[width=0.33\textwidth]{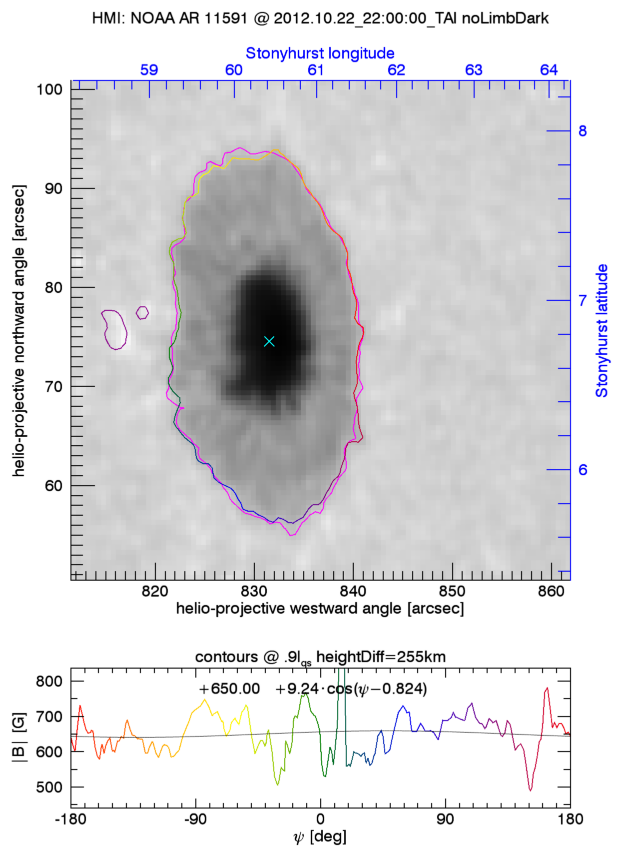}%
\includegraphics[width=0.33\textwidth]{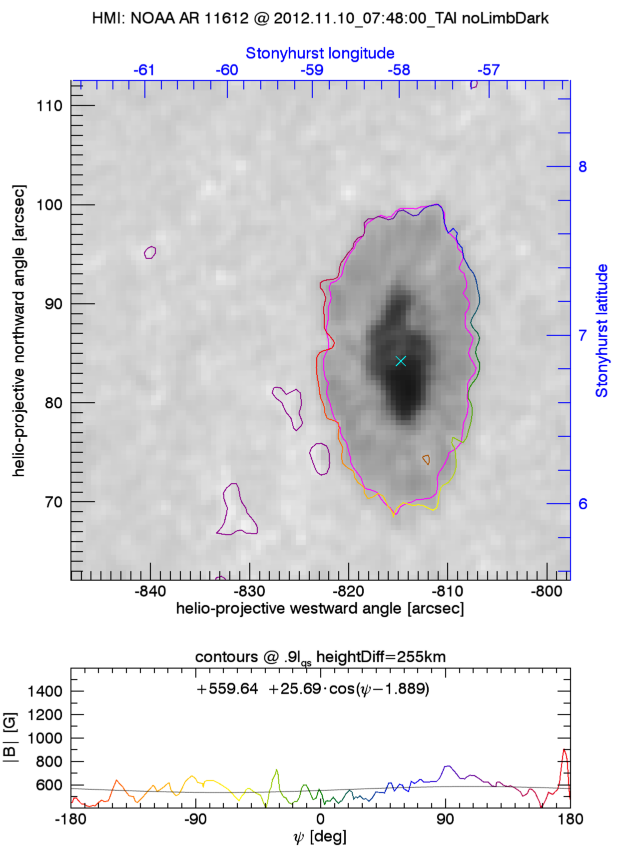}%
\includegraphics[width=0.33\textwidth]{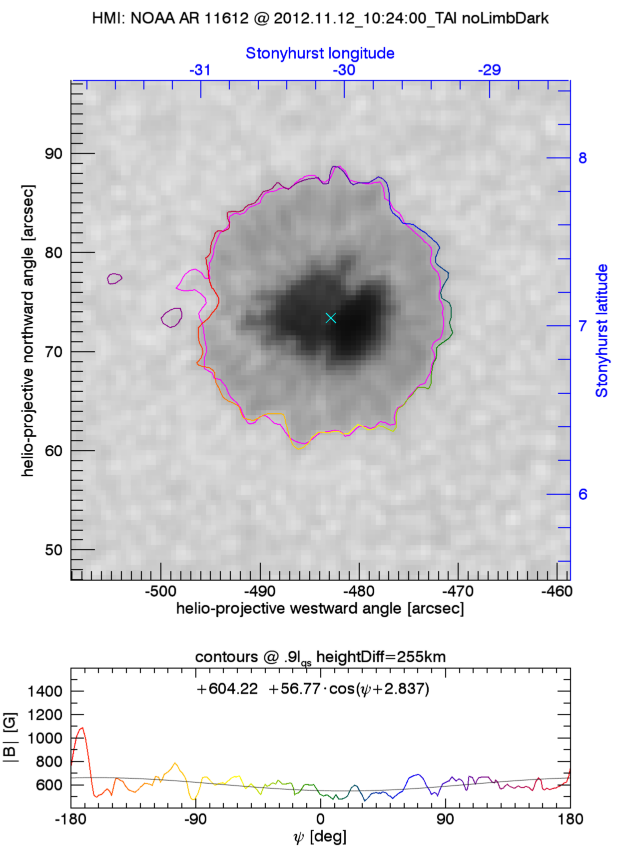}%
\caption{First four panels (from left to right and top to bottom): Same as Fig.~\ref{fig:scr1st0.9h_0}, NOAA\,AR\,11591 for Stonyhurst longitude of the centroid (during the first disc passage) of $-60\degr$, $-30\degr$, $30\degr$ \& $60\degr$. Last two panels: Same but for the second disc passage (NOAA AR 11612) at longitudes of the centroid around $-60\degr$ \& $-30\degr$.
\hfill All panels are screenshots of the movies corresponding to Figs.~\ref{fig:scr1st0.9h_0} \& \ref{fig:scr2nd0.9h_all}, available at \href{https://www.aanda.org/10.1051/0004-6361/202558127/olm}{\texttt{https://www.aanda.org/}}}
\label{fig:scr1st0.9h_rest}
\end{figure*}

\end{appendix}

\end{document}